\documentclass[twocolumn,preprintnumbers,superscriptaddress,amsmath,amssymb,floatfix]{revtex4}

\usepackage{graphicx}
\usepackage{epsfig}
\usepackage{epstopdf}
\usepackage{longtable}
\usepackage{amssymb,amsmath,amsthm}
\usepackage{color}
 
\begin{document}

\title{Determination of spin and orbital magnetization in the ferromagnetic superconductor UCoGe}

\author{M.~W. Butchers}
\affiliation{Department of Physics, University of Warwick, Coventry CV4
7AL, United Kingdom}
\author{J.~A. Duffy}
\email{j.a.duffy@warwick.ac.uk}
\affiliation{Department of Physics, University of Warwick, Coventry CV4
7AL, United Kingdom}
\author{J.~W. Taylor} 
\affiliation{DMSC - European Spallation Source, Universitetsparken 1, Copenhagen 2100, Denmark}
\author{S.~R. Giblin} 
\affiliation{School of Physics and Astronomy, Cardiff University, Cardiff, CF24 3AA, United Kingdom}
\author{S.~B. Dugdale}
\affiliation{H.~H.~Wills Physics Laboratory, University of Bristol, Tyndall Avenue, Bristol BS8 1TL, United Kingdom}
\author{C. Stock}
\affiliation{School of Physics and Astronomy, University of Edinburgh, Edinburgh EH9 3JZ, UK}
\author{P.~H. Tobash}
\affiliation{Los Alamos National Laboratory, Los Alamos, New Mexico 87545, USA}
\author{E.~D. Bauer}
\affiliation{Los Alamos National Laboratory, Los Alamos, New Mexico 87545, USA}
\author{C. Paulsen}
\affiliation{Institut N\'eel, CNRS \& Universit\'e Joseph Fourier, BP 166, 38042 Grenoble Cedex 9, France}

\begin{abstract}
The magnetism in the ferromagnetic superconductor UCoGe has been studied using a combination of magnetic Compton scattering, bulk magnetization, X-ray magnetic circular dichroism and electronic structure calculations, in order to determine the spin and orbital moments. The experimentally observed total spin moment, $M_s$, was found to be -0.24 $\pm$ 0.05~$\mu_B$ at 5~T.  By comparison with the total moment of 0.16 $\pm$ 0.01~$\mu_B$, the orbital moment, $M_l$, was determined to be 0.40 $\pm$ 0.05~$\mu_B$. The U and Co spin moments were determined to be antiparallel.  We find that the U 5\textit{f} electrons carry a spin moment of U$_s \approx$ -0.30~$\mu_B$ and that there is a Co spin moment of Co$_s \approx$ 0.06~$\mu_B$ induced via hybridization.  The ratio U$_l/$U$_s$, of $-1.3 \pm 0.3$, shows the U moment to be itinerant. In order to ensure an accurate description of the properties of 5\textit{f} systems, and to provide a critical test of the theoretical approaches, it is clearly necessary to obtain experimental data for both the spin and orbital moments, rather than just the total magnetic moment. This can be achieved simply by measuring the spin moment with magnetic Compton scattering and comparing this to the total moment from bulk magnetization.

\end{abstract}

\maketitle

UCoGe is one of a family of uranium compounds in which superconductivity and ferromagnetism co-exist.  This unconventional superconductivity was first observed under high pressure in UGe$_2$ \cite{saxena2000}, and more recently at ambient pressure in URhGe \cite{aoki2001} and UCoGe \cite{huy2007}. Unlike conventional superconductivity, in these ferromagnetic superconductors spin-triplet pairing is responsible,  involving electrons with parallel spins. This means that ferromagnetic order is not antagonistic to the superconducting state, and indeed the pairing mechanism is considered to be mediated via ferromagnetic fluctuations.

In UCoGe ferromagnetism and superconductivity have been shown to co-exist using microscopic probes such as muon spin relaxation~\cite{visser2009} and nuclear magnetic resonance \cite{ohta2010}. It is considered to be a weak itinerant ferromagnet, with $T_C \approx 2.4~$K and an ordered magnetic moment between 0.07~$\mu_B$ to 0.18~$\mu_B$.  The superconducting phase occurs below $\approx$ 0.5 K. When the superconducting transition is probed as a function of pressure it is clear that superconductivity also occurs in the paramagnetic phase and the transition extrapolates to a ferromagnetic quantum critical point at the critical pressure~\cite{slooten2009}. Fundamental thermodynamic properties such as magnetization \cite{huy2008,paulsen2012} and superconductivity \cite{huy2008,aoki2009,Ihara2011} are highly anisotropic and numerous experiments have shown the existence of the critical ferromagnetic fluctuations \cite{Ihara2011,stock2011,hattori2012} thought to be necessary for the spin-triplet pairing. 

There has been considerable impetus to understand the electronic structure and magnetism in 5\textit{f} materials, including this series of superconducting ferromagnets, owing to the wide variety of ground state properties exhibited. Theoretical models are required to explain the properties of interactions and fluctuations, and a consequence of this is the need of direct knowledge of the spin and orbital moments. A unique situation can be formed where the spin orbit coupling is typically of a similar magnitude to the crystal field.  The delicate balance between these can lead to different ground states in apparently similar compounds, depending on the degree of localization of the 5\textit{f} electrons. For U, Hund's rules, which describe a local moment system, can be used to obtain the ratio of the orbital moment (U$_l$) and the spin moment (U$_s$). In a free ion the ratio is given by U$_l/$U$_s= -3.29$ for U$^{4+}$ and U$_l/$U$_s= -2.56$ for U$^{3+}$, and values below these are then used to characterize the itinerancy of the 5\textit{f} electrons~\cite{lander1991}.


In UGe$_2$ and UCoGe, defining a single parameter to characterize the degree of itinerancy is insufficient: it has been proposed that the 5\textit{f} electrons simultaneously display both itinerant and localized behavior \cite{mineev2013, chubukov2014}.  In the case of UGe$_2$, there is indeed significant evidence for this so-called electronic duality.  The magnetic order is well described by localized electrons  and analysis of the magnetization was found to be consistent with U$^{4+}$  \cite{troc2012}.  It should be noted, however, that although polarized neutron diffraction (PND) \cite{kernavanios2001}  experiments revealed no evidence of any diffuse magnetization, the orbital to spin moment is reduced with respect to the free-atom value.  However, the muon spin relaxation data also exhibit signatures of the presence of itinerant electrons, with a contribution to the moment estimated to be ~$0.02 ~\mu{_B}$.   The magnetoresistance and specific heat data also have the characteristics expected of itinerant electrons \cite{yaouanc2002,sakarya2010}.

There have been several theoretical studies of the electronic structure and magnetism in UCoGe.  These predict significant spin and orbital U 5\textit{f} magnetic moments, of similar magnitudes, resulting in near cancellation of the total moment.  They all also predict a Co spin moment.  In the case of Refs. \onlinecite{mora2008,czekala2010}, this is parallel to the net U moment, but antiparallel for Ref. \onlinecite{divis2008}.   However, when discussing the underlying electronic structure, it is vital to consider that the Co moment is in all cases \textit{antiparallel} to the U spin moment (and parallel to the U orbital moment).  The apparent flipping with respect to the U total moment arises simply because the U moment is taken to be parallel to whichever is larger out of its spin and orbital contributions: in Refs. \onlinecite{mora2008,czekala2010}, U$_l \ge $U$_s$, but in the calculations of Ref. \onlinecite{divis2008}, U$_l \le $U$_s$.  All these calculations predict a much larger total magnetic moment than is measured experimentally.  To explain this discrepancy  Divi\v{s} \cite{divis2008} suggested the Co moments are not collinear, giving rise to a smaller net moment, as observed in UNiAl \cite{prokes1997}, however, the degree of canting required would have to be $\approx 20^o$ and seems unlikely to be the case due to the highly anisotropic magnetization measurements.  Alternatively this discrepancy could arise from the reduction of the bulk moments due to the presence of strong magnetic fluctuations associated with the proximity of the ferromagnetic critical point.   Furthermore, in contradiction to the theoretical predictions, analysis of experimental PND data suggested that the U and Co spin moments are infact parallel \cite{prokes2010}. This will be discussed in the light of our results later in the paper.

In this Letter we report on our work combining magnetic Compton scattering, magnetization, X-ray magnetic circular dichroism (XMCD) experiments and \textit{ab initio} electronic structure calculations and are hence able to resolve the ground state magnetic configuration of UCoGe. We have been able to determine the site specific spin and orbital contributions to the magnetization.  The XMCD measurements confirm there to be a Co spin moment, antiparallel to the U spin moment, at applied fields between $1 - 6$ T.  It is clear that it is important to be able to resolve both spin and orbital moments, rather than just the total moment when addressing the electronic structure of the actinides, and our approach is ideal for such studies.

UCoGe belongs to the family of ternary compounds U\textit{TX}, with \textit{T} a transition metal and \textit{X} a \textit{p}-electron atom. It crystallizes into the orthorhombic \textit{Pnma} space group with lattice parameters \textit{a} = 6.852 \AA, \textit{b} = 4.208 \AA, and \textit{c} = 7.226 \AA~\cite{canepa1996}, all atoms occupying the same 4\textit{c} symmetry site.  The U atoms arrange themselves in zig-zag chains along the \textit{a}-axis (Fig. \ref{fig1a}a) and each U atom has only two U nearest neighbors at a distance of 0.35 nm characteristic of the critical region between localized and itinerant 5\textit{f}-electron behavior (Hill limit) \cite{hill1970}. The degree of 5\textit{f} localization is down to two things: the direct overlap of corresponding 5\textit{f} wavefunctions on neighboring atoms governed by the Hill limits and also on the 5\textit{f}-6\textit{d} hybridization with ligand states.

\begin{figure}
\centering
\includegraphics[width=0.9\linewidth, viewport=1in 0in 11in 8in]{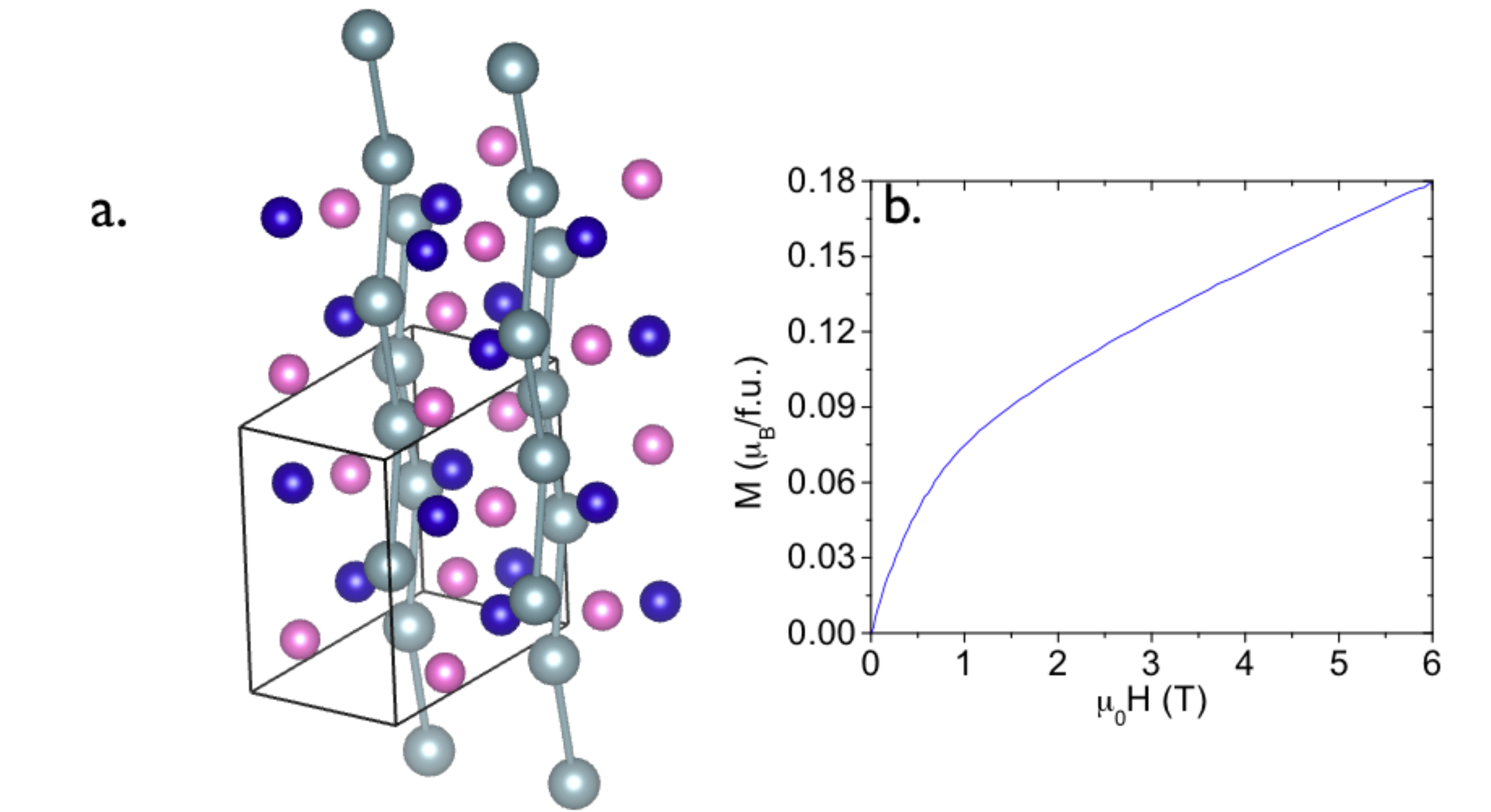}
\caption{\label{fig1a}Color online: a. The crystal structure of UCoGe. The large grey atoms are U showing the zig-zag alignment of the U atoms, dark blue are Co and purple are Ge. The box outlines the unit cell. b. Bulk magnetization of UCoGe at 1.8 K along the c-axis, measured with a SQUID. This includes the total contribution from both the spin and orbital moments. }
\label{fig1a}
\end{figure}

The sample used in this experiment was cut from the sample used in Ref.~\onlinecite{stock2011}, with a mass of 1.08 g. That single crystal was  grown by
the Czochralski technique followed by a pre-defined annealing procedure \cite{huy2009_1}. A small piece was tested and found to have a residual resistance ratio (RRR) of 4. The bulk sample used for magnetic Compton and XMCD was characterized using d.c. magnetization and a.c. susceptibility. From Arrott plots it was found the sample has a ferromagnetic transition of 2.4~K and the onset of the superconducting transition is seen at 0.6~K. The extrapolated value of the upper critical field obtained from a.c. susceptibility measurements with the field applied along the $c$ axis are coincident with previous work on a sample with a RRR of 30~\cite{deguchi2010}, demonstrating that even for such a large sample the fundamental properties of UCoGe remain. Indeed the anisotropic fluctuations seen in this large sample~\cite{stock2011} are also observed in much smaller samples~\cite{hattori2012}.

In a Compton scattering experiment, the 1D projection of the electron momentum density
distribution is obtained via measurement of the energy distribution of high energy X-rays scattered from the sample being studied. A monochromatic X-ray beam is used, and at a defined scattering angle the scattered photons have an energy spectrum that is directly related to the sample's electron momentum distribution via the Klein-Nishina cross-section \cite{klein1929}.   The Compton profile is defined as a 1D projection (onto the scattering vector) of
the electron momentum distribution, $n(\textbf{p})$ \cite{dugdale2014}, where the $\textit{z}$ direction
is taken parallel to the scattering vector: 

\begin{equation} J\left( {p_{z}}
\right)\, = \,\int\!\!\!\int {n\left( {\textbf{p}} \right)} \,dp_{x} dp_{y}.
\label{eq1}
\end{equation}

If the incident beam has a component of circular
polarization, the scattering cross-section contains a term which is spin dependent
\cite{bell96}.  In principle, the spin dependence may be isolated by either flipping the direction
of magnetization or the photon helicity parallel and antiparallel with respect to
the scattering vector. Either method results in a $\textit{magnetic}$ Compton
profile (MCP), $J_{mag}(p_{z})$, that is only sensitive to the net spin moment of
the sample, and is defined as the 1D projection of the spin-polarized electron
momentum density:

\begin{equation}
 J_{mag} \left( {p_{z}}  \right)\, = \,\int\!\!\!\int {\left[ {}
\right.n^{ \uparrow} \left( {\textbf{p}} \right)} \, - \,n^{ \downarrow} \left(
{\textbf{p}} \right)\left. {} \right]\,dp_{x} dp_{y}.
\label{eq2}
\end{equation}

Here $n^{\uparrow} ($\textbf{p}$)$ and $n^{\downarrow
}($\textbf{p}$)$ are the momentum densities of the majority and minority spin
bands.  The integrated area of this magnetic Compton
profile (MCP) provides the total spin moment per
formula unit of the sample.   The orbital moment is not observed \cite{cooper93}, and its value can be determined
simply by comparison with a bulk magnetization measurement. Since the MCP is the difference between two measured Compton profiles,
components arising from spin-paired electrons cancel, as do most sources of
systematic error.  The high X-ray energies
used in the experiments, typically $100-200$keV, mean that the bulk
electronic structure is measured.  Crucially the incoherent nature of Compton scattering means
that all local and itinerant contributions to the spin moment are observed (for example see Refs \cite{duffy1998, duffy2000}). 

In an experimental measurement, the scattering signal obtained is proportional to the Compton profiles defined in Eqs. 1 and 2. The spin moment may then be determined using the flipping ratio, $R$, of the integrated magnetic and charge measurements, where

\begin{equation}
R \propto \frac{\int J_{mag} {\left( p_{z} \right)} dp_z}{\int J {\left(
p_{z} \right)} dp_z}.
\label{eq3}
\end{equation}

The spin moment can be obtained quantitatively from the experimental data in a straightforward manner, given that it is proportional to the measured flipping ratio (for example see Ref. \cite{duffy2010}).  It is determined via comparison with a reference measurement, made in the same experimental set up, of the flipping ratio for a sample with a known spin moment.  In our experiment, we used Ni, for which the spin moment (0.56~$\mu_B$) is well established.  

The spin-polarized Compton profiles presented in this Letter were
measured on beam line ID15 at the ESRF.  An Oxford Instruments Spectromag cryomagnet was used to obtain fields of
5 T and maintain a sample temperature of 1.5 K.  The energy spectrum of the scattered
flux was measured using a 13-element Ge detector at a mean scattering angle of
172$^{\circ}$. The incident energy of 220~keV and
scattering angle of 172$^{\circ}$ resulted in a resolution of 0.44 a.u. of momentum (where
1 a.u.=$1.99\times10^{-24}$\thinspace kg\thinspace m\thinspace s$^{-1}$).  The magnetic signal was isolated by flipping the magnetic field applied to the sample.  The data were corrected for energy-dependent detector efficiency, sample absorption, and the relativistic scattering cross-section. The XMCD experiment was performed on I06 at Diamond Light Source. The vector superconducting magnet on the I06 branchline has the capability of providing a sample environment down to 1.4~K in a magnetic field of 6~T. The branchline is fed by an APPLE-II undulator with a useful energy range (i.e. that which is circularly polarized) between 100 - 1300 eV. All XMCD measurements were performed with a fluorescence detector.

\begin{figure}
\centering
\includegraphics[width=0.85\linewidth,clip=true]{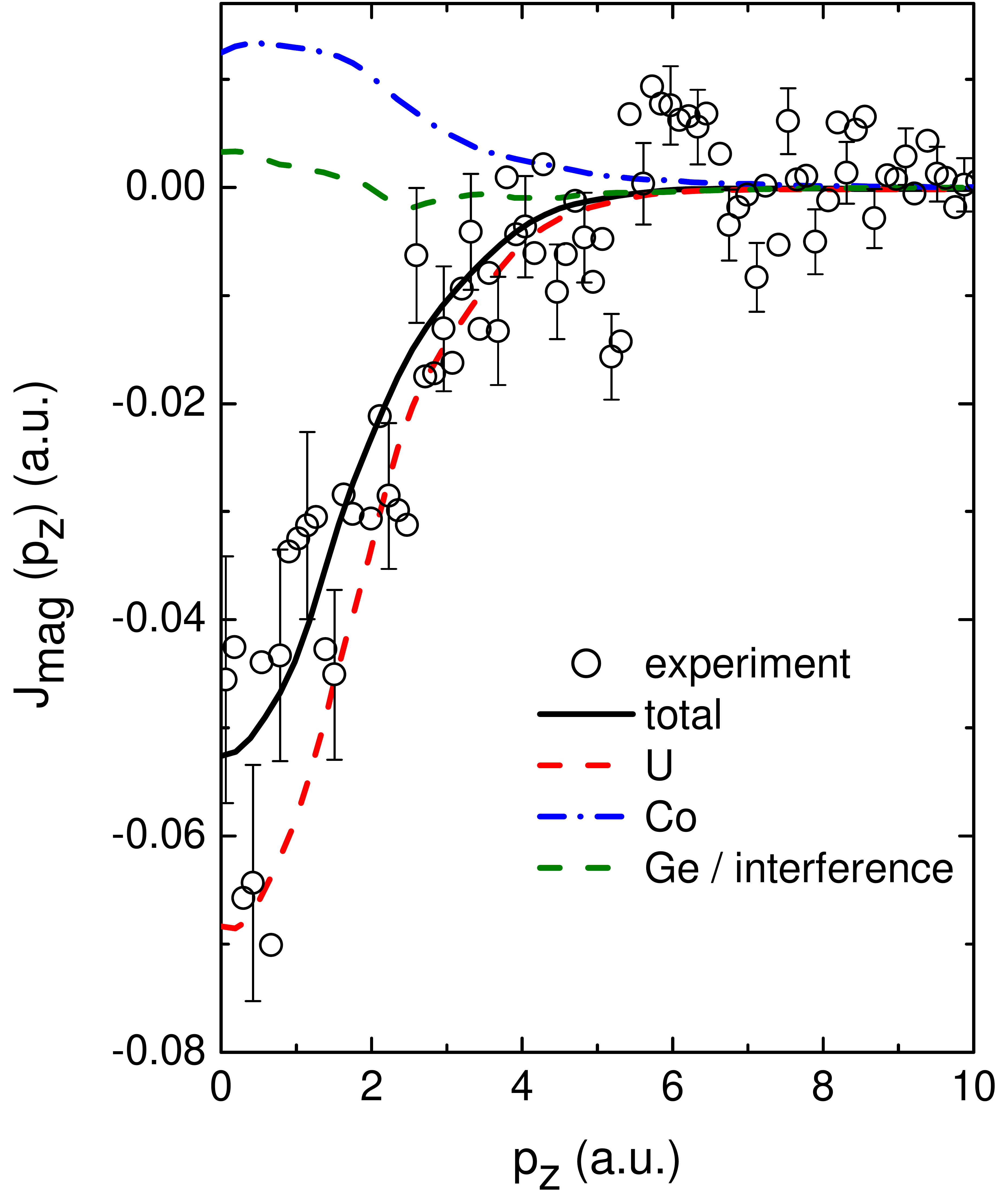}
\caption{Color online: Magnetic Compton profile of UCoGe along the c-axis taken at 5 T, shown with spin density predicted by KKR calculations normalized to the spin moment of -0.24 $\pm$ 0.05~$\mu_B$ and decomposed into the site projected profiles.}
\label{fig1}
\end{figure}

The MCP of UCoGe measured in a field of 5 T at 1.5 K is shown in Fig.~\ref{fig1}. The total spin moment, $M_s$, in this UCoGe sample was determined to be -0.24 $\pm$ 0.05~$\mu_B$. Using a direct comparison of the bulk magnetization which is shown in Fig.~\ref{fig1a}b and gives the measured total magnetic moment as 0.16 $\pm$ 0.01~$\mu_B$, the orbital moment is then determined to be 0.40 $\pm$ 0.05~$\mu_B$. We note that the magnetisation data were obtained at 1.8 K, but little change is expected at 1.5 K in our 5 T applied field ~\cite{troc2010} and would not affect our orbital moment value. The contribution to the MCP from electrons associated with specific atoms are generally experimentally distinguishable allowing the identification of site specific moments. However, the electron momentum distribution of U 5\textit{f} and Co 3\textit{d} are essentially (within experimental error) indistinguishable.

To separate the contribution of site specific U and Co moments, electronic structure calculations have been performed in the local spin density approximation (LSDA) using the SPR-KKR package \cite{ebert2011}.  The electronic structure and magnetic moments obtained from the calculations are consistent with previous results \cite{sam}. The spin-resolved electron momentum density, and hence the MCP, can be calculated directly from the electronic structure, enabling comparison with our experimental  profile. Using the theoretical and experimental profiles together can give detailed information about the underlying electronic structure and magnetic moments \cite{dixon1998, utfeld2009, haynes2012}. The total spin and orbital moments obtained ($-0.71~\mu_B$ and $1.21~\mu_B$ respectively) from the calculation are both a factor of three larger than the experimental values. The calculated spin moment has been scaled to the experimental value, as the LSDA calculations do not take into account spin fluctuations \cite{aguayo2004} which are expected to reduce the moment \cite{haynes2012} and the resultant fit of the calculation to the MCP is shown in Fig. \ref{fig1}.  Scaling the contributing moments by the same proportion suggests that the U 5\textit{f} electrons carry a spin moment of U${_s} \approx$ -0.30 $\mu_B$ and that there is a Co spin moment of Co${_s} \approx 0.06~\mu_B$. From this, the U orbit/spin ratio is deduced to be $-1.3 \pm\ 0.3$, showing that the U 5\textit{f} electrons are highly itinerant in UCoGe.  In this analysis it is assumed that the predicted individual U and Co moment contributions scale by the same proportion as the total moments: this seems plausible, given that the total spin and orbital moments both scale by the same factor.  We note that in our previous work on the NbFe$_2$ system \cite{haynes2012}, where spin fluctuations are also thought to be responsible for the reduced experimental spin moment, we were able to demonstrate that the different spin contributions did follow the total moment and that the electronic structure appeared to be unaffected. Even if we drop this assumption, then because we know there is a non-zero Co spin moment which is antiparallel to the U spin moment (from XMCD data, which are discussed below), the U$_{s}$ value must be greater than $-0.24~\mu_B$, which means that the  U orbit/spin ratio is certainly less than $-1.6 \pm\ 0.3$.  This low value, when compared to the free atom value, suggests strong 5\textit{f} - 5\textit{f} overlap and strong 5\textit{f} - 3\textit{d} hybridization. It is also shown that the Co spin moment obtained from MCS is a moderately large 0.06 $\mu_B$ induced by strong hybridization with the U, suggesting that the Co orbitals play a significant role in the delocalization of the U electrons. 

\begin{figure}
\centering
\includegraphics[width=0.9\linewidth,clip=true]{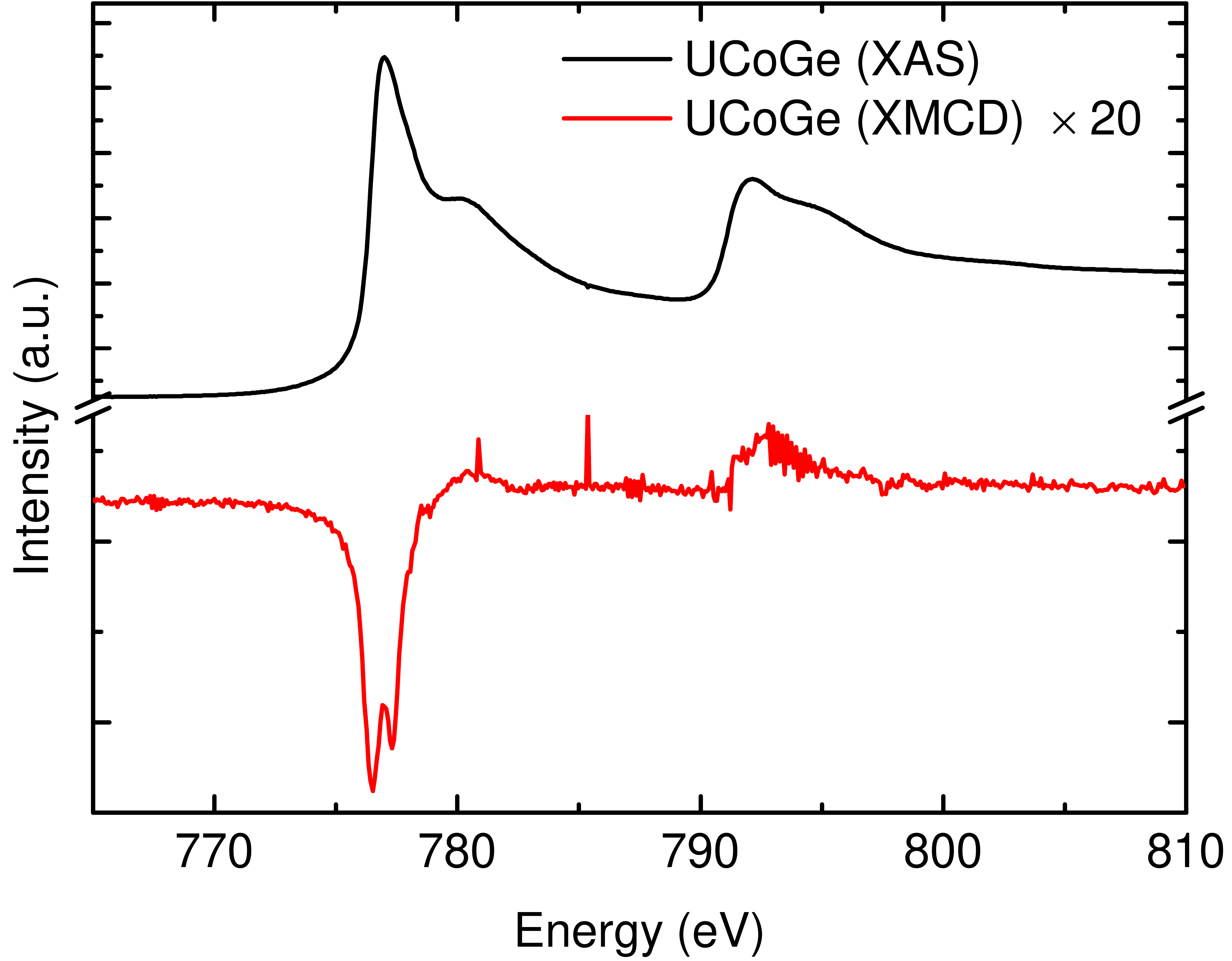}
\caption{\label{xmcd} Color online: XAS and XMCD signal of UCoGe at 1.5~K and 6~T showing the Co dichroic signal.}
\end{figure}

To confirm the antiparallel alignment of the U and Co moment we have used XMCD obtained from  absorption spectroscopy (XAS)  to study the magnetization at specific elemental edges.  Fig. \ref{xmcd} shows a typical XAS and XMCD signal below the ferromagnetic transition where dichroism at the Co L$_3$ and L$_2$ edge was observed. Using EuCoO$_3$ as a standard Co reference ~\cite{burnus}, the valence state of UCoGe is in the Co$^{3+}$ state and the Co spin moment aligns with the field, as the dichroism is the same sign for Co where it is assumed the Co moment aligns with the field. This confirms the result of the antiparallel alignment observed with MCP.  One complication of investigating UCoGe with XMCD is that the positions of the Co L$_3$ and the U N$_4$ edge overlap strongly. Indeed the difference in binding energy is only 0.2 eV. However, we do not anticipate a significant dichroic signal at the U N$_4$ edge since any dichroism at the U N$_5$-edge was too small to be observed in our experiment. (It is important to note that the dichroism from the N-edge may be an order of magnitude smaller than that from M-edge transitions~\cite{94}).  We have not attempted a quantitative analysis from the XMCD because of the overlap. A very recent study of XMCD at the U M edges complements our experimental observations~\cite{taupin2015}. For an applied field of 5 T, their orbital moment ($\approx 0.3~\mu_B$) is similar to ours, but their spin moment ($\approx -0.14~\mu_B$) is smaller. This then leads to a higher U orbit/spin ratio of $\approx -2.3 \pm\ 0.3$. The origin of this discrepancy is not clear. However, it is possible that it arises from the analysis required to obtain the spin moment from XMCD, which is more difficult in $5f$  systems than for the orbital moment (for example see \cite{wilhelm2013}). As discussed above, our U spin moment is greater than $-0.24~\mu_B$, and hence our orbit/spin ratio is less than  $-1.7 \pm\ 0.3$.

In order to make progress with the theoretical description of these U based superconductors, experimental measurements of spin and orbital magnetic moments are required.  Bulk magnetization studies provide a first step, but obviously only permit measurement of the total magnetic moment.  For materials where the total moments are small but arise from the cancellation of the spin and orbital moments, a measurement providing their individual contributions is crucial.  A recent study using PND was published, with a number of significant findings reported~\cite{prokes2010}.  Firstly, in contrast with the various theoretical studies, the authors' analysis determined the U 5\textit{f} and Co 3\textit{d} spin moments to be aligned in parallel, rather than antiparallel.  Secondly, the relative contributions to the magnetization density changed as the applied magnetic field was increased, with the Co spin moment being enhanced relative to the U moment at 12~T compared to 3~T.  Taking their derived U spin and orbital moments gives orbital/spin ratios of $-3.6 \pm 1.5$ (3~T) and $-2.9 \pm 1.6$ (3~T), which are somewhat larger than our value.  However, the total magnetic moments determined from the PND data were significantly less than the total bulk magnetization.  This discrepancy was ascribed to the existence of an itinerant moment which could not be attributed to either the U or Co sites.

In summary we have used magnetic Compton scattering, XMCD and magnetization measurements to characterize a bulk sample of UCoGe, and to clarify the properties of the site specific moments. It has been shown clearly that the U and Co moment are aligned antiparallel. Moreover, UCoGe is not composed of two large opposing orbital and spin moments, but instead consists of two opposing fairly weak spin and orbital moments. The magnitude of the individual moments are indicative of a strongly delocalized electron system, with the delocalization mechanism being a strong overlap between U 5\textit{f} and Co 3\textit{d} electrons which consequently result in a non-negligible Co moment. By use of XMCD experiments the alignment of the moments are determined to be in agreement with \textit{ab initio} calculations, but in contrast with PND measurements. Most of the total moment comes from the orbital contribution, but the majority of the spin moment comes from the U 5\textit{f} electrons with a small non-negligible contribution from the Co 3\textit{d} electrons. Magnetic Compton scattering in combination with standard magnetization is a powerful probe to separate spin and orbital moments, and could be pertinent to the iridate pyrochlore systems. This work highlights the importance of determining not only the total moment but also the spin and orbital contributions.

\section*{Acknowledgments}
We acknowledge UK EPSRC via grant EP/F021518. S.R.G. acknowledges the European Community Ñ Research Infrastructures under the FP7 Capacities Specific Programme, MICROKELVIN, 228464. We are grateful for the support of Veijo Honkim\"{a}ki and the staff of ID15 at the ESRF. We thank Diamond Light Source for access to beamline I06 (SI-7010) and for the support of Sarnjeet Dhesi and the staff. Work at Los Alamos National Laboratory was performed under the U.S. DOE, OBES, Division of Materials Sciences and Engineering.

\end{document}